# Data-driven nonparametric Li-ion battery ageing model aiming at learning from real operation data – Part A: Storage operation

M. Lucu[a,b,*], E. Martinez-Laserna[a], I. Gandiaga[a], K. Liu[c], H. Camblong[b,d], W.D. Widanage[c], J. Marco[c]

[a] Ikerlan Technology Research Centre, Basque Research and Technology Alliance (BRTA). P° J.M. Arizmendiarrieta, 2, 20500 Arrasate-Mondragón, Spain
[b] Department of Systems Engineering & Control, University of the Basque Country (UPV/EHU), Europa Plaza, 1. 20018 Donostia-San Sebastian, Spain
[c] WMG, University of Warwick, Coventry CV4 7AL, UK
[d] ESTIA Research, Ecole Supérieur des Technologies Industrielles Avancées (ESTIA), Technopole Izarbel, 64210 Bidart, France

ABSTRACT

Conventional Li-ion battery ageing models, such as electrochemical, semi-empirical and empirical models, require a significant amount of time and experimental resources to provide accurate predictions under realistic operating conditions. At the same time, there is significant interest from industry in the introduction of new data collection telemetry technology. This implies the forthcoming availability of a significant amount of real-world battery operation data. In this context, the development of ageing models able to learn from in-field battery operation data is an interesting solution to mitigate the need for exhaustive laboratory testing.

In a series of two papers, a data-driven ageing model is developed for Li-ion batteries under the Gaussian Process framework. A special emphasis is placed on illustrating the ability of the Gaussian Process model to learn from new data observations, providing more accurate and confident predictions, and extending the operating window of the model.

This first paper focusses on the systematic modelling and experimental verification of cell degradation through calendar ageing. A specific covariance function is composed, tailored for use in a battery ageing application. Over an extensive dataset involving 32 cells tested during more than three years, different training possibilities are contemplated in order to quantify the minimal number of laboratory tests required for the design of an accurate ageing model. A model trained with only 18 tested cells achieves an overall mean-absolute-error of 0.53% in the capacity curves prediction, after being validated under a broad window of both dynamic and static temperature and SOC storage conditions.

## 1. Introduction

Lithium-ion (Li-ion) battery technology has gained a significant market share as the principal energy storage solution for many industrial applications, mainly due to its high energy efficiency and high specific energy and power [1,2]. However, Li-ion batteries are still relatively expensive compared to other storage technologies, and their performance is known to decline over time and use, which threatens their competitiveness against more affordable solutions [2,3]. In order to overcome such barriers, the global research in Li-ion batteries focusses on different paths. On the one hand, the next generation battery technology is wanted to be developed working on improved or new materials, in order to increase the specific energy and energy density [4,5], minimise side reactions [6], improve safety [7] and reduce material costs [8]. On the other hand, optimised sizing of the storage systems [9], second-life business strategies [10] and the design of effective operation strategies for the currently commercialised Li-ion battery technologies allow the reduction of the total cost of ownership, making profitable the implementation of large-scale Li-ion energy storage systems [11]. The latter points are strongly conditioned by the development of accurate battery ageing models. In fact, accurate ageing predictions could help to identify too heavy battery operating conditions and avoid the need for system replacement. Inversely, scenarios of more intensive use of the battery could be contemplated to increase the profitability of the application [12].

Different forms of ageing models have been widely proposed in the

* Corresponding author at: Ikerlan Technology Research Centre, Basque Research and Technology Alliance (BRTA). P° J.M. Arizmendiarrieta, 2, 20500 Arrasate-Mondragón, Spain.

*E-mail address:* mlucu@ikerlan.es (M. Lucu).

literature, with varying levels of complexity, accuracy and representativeness of the internal physics and chemical processes in the battery [13]. Electrochemical models are known to provide a good mathematical representation of the internal variables of the battery, such as the thickness and conductivity of the Solid Electrolyte Interface (SEI) [14,15]. However, such a detailed mathematical representation implies increased levels of complexity and computational cost. Moreover, the development of electrochemical models supposes an extensive parametrisation phase typically requiring cell disassembly [13]. Models based on in-field measurable variables are argued to be more suitable for implementation in real-world applications [16]. Empirical models rely on experimental ageing tests while semi-empirical ageing models add a physicochemical support to the mathematical empirical data fitting phase [17,18]. Developing such ageing models generally consists of capturing the relations between battery's health indicators (e.g. capacity or internal resistance [19]), and stress-factors; the most widely used factors cited in the literature include operating time, temperature, State of Charge (SOC), Ah-throughput, C-rate and depth of discharge [16].

A significant challenge for the development of such conventional ageing models is the amount of laboratory tests required to verify the accuracy of the model under realistic operating conditions. Conventional models are typically parametrised using laboratory tests carried out at constant ageing conditions [20,21]. Furthermore, extensive validation procedures involving constant ageing conditions, slowly varying dynamic conditions and realistic ageing profiles are recommended to surround accurate lifetime predictions in a context of real-world operation [16]. However, even such a time and cost-intensive validation procedure cannot ensure accurate predictions for a large diversity of dynamic or realistic profiles, particularly when taking into account the reported path dependence within many battery ageing factors [22].

As suggested in a previous publication, a suitable solution to reduce the number of laboratory tests could be the development of ageing models capable to continuously learn from streaming data [23]. Following this approach, reduced laboratory tests could be used to develop a preliminary ageing model. Further, once the battery pack has been implemented and deployed, in-field data extracted by the data acquisition system could allow updating the preliminary ageing model. In this way, the ageing model would be continuously upgraded, improving prediction accuracy, extending the operating window of the model itself and providing useful information for predictive maintenance, adaptive energy management strategies or business case redefinition.

In a previous study, a critical review on self-adaptive ageing models for Li-ion batteries was presented, in which the Gaussian Process (GP) method was identified as the most promising candidate [23]. In fact, beyond their ability to perform probabilistic, relatively robust and computationally acceptable predictions, these models enjoy the very interesting advantage of being nonparametric: in other words, the complexity of these models depends on the volume of training data. Within the context of Li-ion battery ageing prediction, this implies:

- *A progressive spread of the operating window for the model.* Each time a new data sample related to previously unobserved operating conditions is included within the training set, additional knowledge is obtained about the influence of stress-factors on ageing. The resulting models should provide an increasingly comprehensive picture of the ageing of Li-ion batteries.
- *A higher level of specialisation of the model.* The preliminary ageing model developed from the laboratory ageing data could be upgraded by including new training data extracted from the in-field operation. In-field data encodes the intrinsic operating profiles of each application, as well as the corresponding battery ageing. This implies the possibility to move from a generic ageing model to a specialised model tailored to the specific applications.

Each time input values are presented to the model to perform a prediction, the GP model retrieves similar data samples in the training dataset to produce analogous predictions. A continuously fed training dataset implies an increased number of similar data, allowing more accurate and confident predictions.

From a broader perspective, the most critical gaps identified in the literature regarding data-driven Li-ion ageing models are i) the underutilisation of key predictive features (e.g. values of the different stress-factors) and ii) the insufficient validation of the proposed models [23]. These gaps strongly limit the accuracy and applicability of the models within the context of real deployment. In this sense, investigation in data-driven Li-ion ageing models should be more focussed on the implementation or discovery of features presenting strong predictive capabilities (as suggested in [24]), as well as the deeper validation of the developed models under broad operating conditions. Moreover, most of the data-driven ageing models proposed in the literature refers to the degradation of the Li-ion batteries when the cell is electrically cycled. However, some applications are characterised by a dominant storage operation of the battery system (e.g. Uninterruptible Power Supplies, Electric Vehicle applications, etc.), and the development of pure calendar ageing models is also necessary.

The GP framework has already been introduced for Li-ion battery ageing predictions [25-33]. The present study aims to extend existing research by integrating the following main contributions:

(i) The analysis of the ability of GP models to learn from new data, illustrating their capability to provide more accurate and confident ageing predictions when integrating previously unobserved operating conditions, extending this way the operating window of the model.
(ii) The introduction of compositional covariance functions tailored to Li-ion battery ageing prediction. In the literature, the ageing models based on the GP framework are limited to the utilisation of conventional covariance functions, selected to achieve a minimal prediction error for a specific dataset. Nevertheless, the covariance function encodes the basic assumptions of the model about the system under study and have a significant impact on the prediction and learning capabilities of the GP. Therefore, the development of the covariance function should be tailored to the application of Li-ion battery ageing and justified beyond the intrinsic patterns of a specific dataset.

Additionally, this study also explores several points uncovered in the literature, introducing the following secondary contributions:

(i) The quantification of the minimal number of laboratory tests required for the design of an accurate calendar ageing model for a broad operating window.
(ii) The validation of the proposed ageing model with an extensive experimental ageing dataset, involving 30 cells tested during more than three years at static conditions, and 2 additional cells tested at dynamic operating conditions.
(iii) The sensitivity analysis of the capacity loss with respect to the different stress-factors, from the point of view of the developed model. As explained in this paper, the developed covariance function shares the particularity of quantifying the relevance of each input variable for predicting the defined output variable. This could provide some intuitions about e.g. which stress-factors are the most impactful on the capacity loss, producing useful insights for the design of energy management strategies. Such analysis was not performed in the field of Li-ion battery ageing prediction.

The body of the research undertaken is presented through a two-part series. This paper focusses on the systematic modelling and experimental verification of cell degradation through calendar ageing. Conversantly, the second paper [34] addresses the same research challenge when the cell is electrically cycled. During many real-world conditions, the cell will be subject to both calendar and cyclic ageing. The relative importance of each will be highly dependent on the nature of the use-case. The integration of both forms of ageing, within the context of defining a holistic view of lithium-ion degradation modelling is a challenging research task, discussed further within [35,36] and is the subject of ongoing research by the authors further extending the research presented here and in [34].

This paper is structured as follows, Section 2 describes the experimental ageing tests carried out in order to produce the ageing data. The raw data obtained from the experimental tests are analysed and preprocessed before the development of the model. Section 3 details the processing of the raw data and evaluate the relevance of the obtained data for ageing modelling. Section 4 introduces the general background of the GP theory, and Section 5 presents the development of the proposed calendar ageing model under the GP framework. In Section 6 and 7, the prediction results of the developed model are presented for the cells tested at static and dynamic storage conditions, respectively. Furthermore, both sections aim to illustrate the ability of the GP model to learn from new data observation. Section 8 discusses the obtained results, leading to the identification of the limitations of the study and opening the way to further works. Finally, Section 9 closes the study depicting the main conclusions.

## 2. Experimental calendar ageing data

Within the context of the European project titled as Batteries2020, extensive experimental works were carried out over a time span of more than three years, in order to analyse the ageing of Li-ion batteries, covering different possible operations. The capacity retention of a 20 Ah Lithium Nickel-Manganese-Cobalt (NMC 4:4:2) cathode-based pouch cell with a graphite anode was evaluated. The nominal characteristics of the cell, as well as the operating conditions recommended by the manufacturer are specified Table 1.

A testing batch of 124 cells, related to the study of the ageing in cycling operation, is described in the second paper of the series corresponding to the development of a cycle ageing model [34]. In this first paper, the experimental works associated with the study of the calendar operation will be presented.

From the ageing point of view, the operation of a Li-ion battery in storage is conditioned by the level of different stress-factors, mainly identified in the literature as the storage temperature and State-Of-Charge (SOC) [16]. A total of 32 cells were tested in temperature-controlled climatic chambers, at different combinations of such stress-factors. Periodical characterisation tests were carried out at 25 °C in order to evaluate the progressive capacity retention of the cells. The determination of the capacity started 30 min after its surface temperature reached 25 °C degrees, ensuring that the cells has stabilised at the target temperature. The test started with a constant current – constant voltage (CC-CV) charge: the CC charge was done at 6.667 A (C/3) until reaching 4.15 V, and the following CV charge was stopped when achieving current values below 1 A (C/20). After a period of 30 min, the cell was discharged using a CC discharge current at 6.667 A (C/3) until the terminal voltage measured 3 V, followed by a pause period of 30 min. The procedure was repeated three times. The capacity value obtained in the last repetition was considered as the cell capacity.

**Table 1**
Nominal characteristics of the tested cell.

| Electrical characteristics | |
|---|---|
| Nominal voltage [V] | 3.65 |
| Nominal capacity [Ah] | 20 |
| AC impedance (1 kHz) [mOhm] | < 3 |
| Specific energy [Wh.kg-1] | 174 |
| Energy density [Wh.L-1] | 370 |
| **Operating conditions** | |
| End of charge voltage [V] | 4.15 |
| End of discharge voltage [V] | 3.0 |
| Recommended charge current [A] | 10 |
| Maximum discharge current (continuous) [A] | 100 |
| Operating temperature [ °C] | −30/+55 |
| Recommended charge temperature [ °C] | 0/+40 |

**Table 2**
Calendar ageing tests matrix, for the tests at static ageing conditions.

| Temperature [ °C] | SOC [%] | | | | | |
|---|---|---|---|---|---|---|
| | 100 | 80 | 65 | 50 | 35 | 20 |
| 25 | | 3 | | 3 | | |
| 35 | 3 | 3 | 3 | 3 | 3 | 3 |
| 45 | | 3 | | 3 | | |

Depending on the variability of the stress-factors' profiles in the whole duration of the tests, two types of ageing experiments were distinguished, namely i) the ageing tests at static operating conditions and ii) the ageing tests at dynamic operating conditions.

### 2.1. *Experimental ageing tests at static operating conditions*

In the ageing tests performed at static conditions, the value of the stress-factors remained constant throughout the whole duration of the tests. A total of 30 cells were tested at 10 different storage conditions, specified in Table 2. These tests were performed in the laboratories of ISEA-RWTH, which was a partner of the Batteries2020 European project consortium. The cells were characterised approximately every 28 days. In order to ensure the repeatability of the results, 3 cells were allocated to each testing condition. The capacity curves resulting from the experimental ageing tests at static conditions are observable in Fig. A1(a–c), Appendix A. The variability of the capacity curves obtained for each tested storage conditions is indicated in Table A1, Appendix A. As already reported in the literature [37], a clear effect of temperature and SOC levels is observable, as higher temperature and SOC levels are known to induce faster capacity loss.

### 2.2. *Experimental ageing tests at dynamic operating conditions*

As the battery stress conditions in real-world applications are not constant over time, the developed ageing models should be able to perform accurate predictions at dynamic operating profiles. The ability of the GP model to learn from dynamic profiles should also be analysed. Therefore, 2 additional cells were tested in the laboratories of Ikerlan Technology Research Centre, under variable ageing conditions, namely the temperature and SOC level were modified between each periodic characterisation experiment. The cells were characterised approximately every 28 days. The obtained capacity curves and the corresponding dynamic operating profiles are depicted in Fig. A1(d-e), Appendix A. It is noteworthy that the lower capacity measurement observable in Fig. A1(d), between days 1200 and 1300 were induced by environmental testing errors, due to temperature control issues in the climatic chambers.

## 3. Data preprocessing

In the context of data-driven modelling, an important step is to analyse and preprocess the raw data before any modelling task, in order to address data inconsistency and noise issues and achieve effective models [38]. The capacity curves obtained from the experimental ageing tests described in Section 2 could be decomposed into four distinct phases, as illustrated in Fig. 1.

The first phase corresponds to an initial capacity rise appearing at the Beginning Of Life (BOL). This is clearly observable in cells exposed to relatively light ageing conditions (e.g. low temperature and SOC levels, Fig. A1, Appendix A) and matches with the calendar experimental data already published in the literature [40,41]. According to the literature, the capacity recovery could be induced by a slow, compensating flow of active lithium between the passive and the active part of the anode, where the passive part represents the geometric excess of the anode with respect to the cathode [41–43]. However, no clear relationship was found between the initial capacity recovery and any ageing mechanism. Therefore, it was assumed that the initial capacity recovery phenomenon is not provoked by an ageing mechanism itself and does not have any influence on the subsequent ageing trend of the cells. This assumption should be verified in further work (see Section 8). Accordingly, the data corresponding to Phase 1 was discarded for the development of the ageing model. During the data preprocessing stage, the maximal capacity point of each cell was designated as the BOL point and assigned to the 'zero storage days' state.

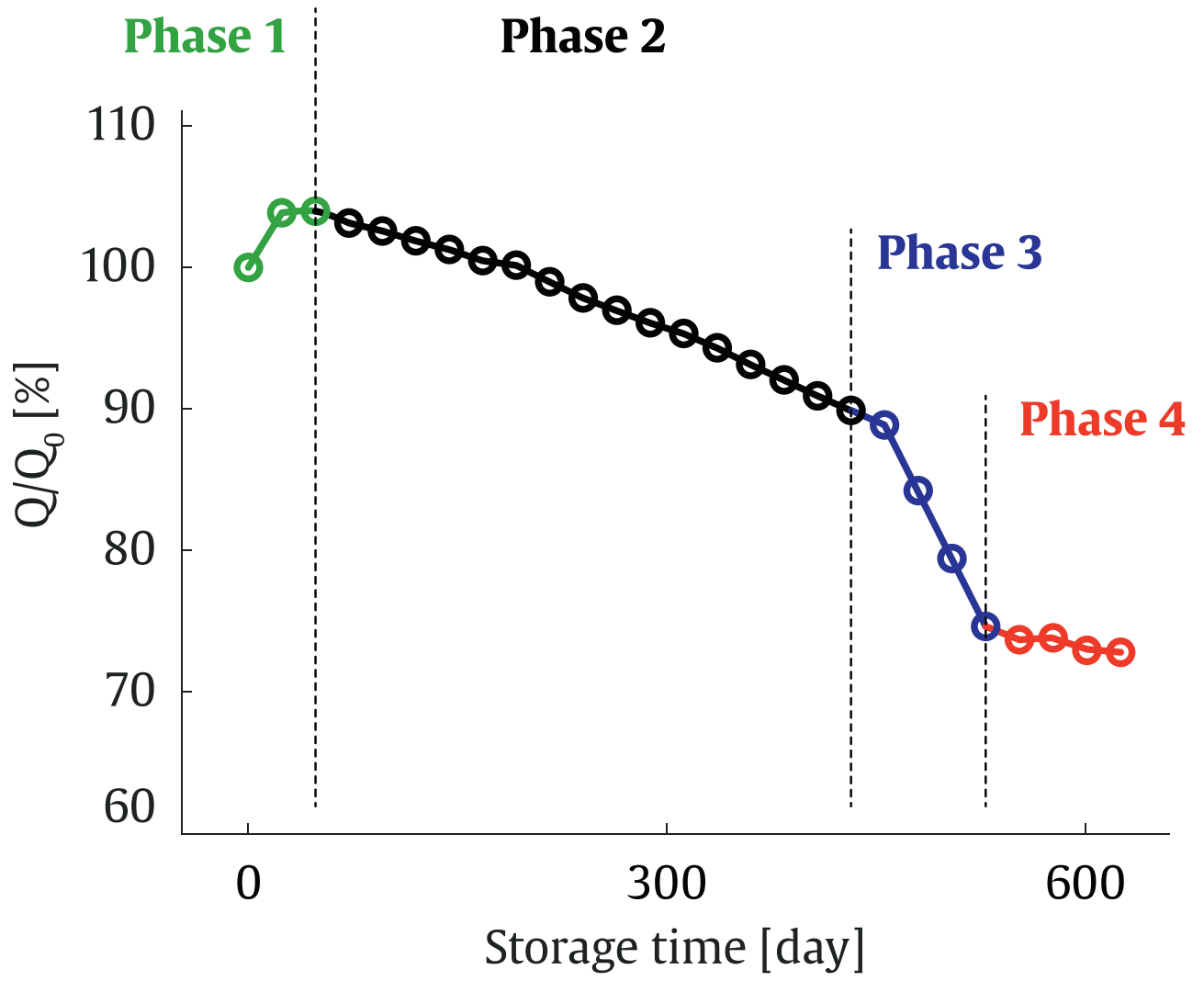

**Fig. 1.** The four different phases of the capacity retention curve of the cells. The first phase is an increase of the capacity, the second is progressive degradation, the third phase is a sudden capacity drop and the fourth phase is a slowdown of the capacity loss. Modified from [39].

After the initial capacity increase, a progressive rate-constant decrease of the cell capacity is observable, identified in Fig. 1 as Phase 2. This phase is the main phase corresponding to the regular degradation of Li-ion batteries, mainly linked to the growth of the SEI, in calendar operation [44].

After a rate-constant decline of capacity, some cells showed a clear acceleration of the ageing rate (Phase 3), especially those stored at high temperature and SOC levels (e.g. black curves, Fig. A1(b), and green and blue curves, Fig. A1(c), Appendix A). Similar behaviour has been reported in the literature for calendar ageing [19,37,45]. One of the cells stored at 35 °C and 100% SOC (black curves, Fig. A1(b), Appendix A) was reserved to carry out post-mortem analysis, and revealed lithium plating at one electrode edge, even after a short storage time of 240 days [39]. The early appearance of lithium deposition for this cell suggests a most advanced propagation of lithium plating for the cells aged at similar and more significant ageing conditions. Therefore, the sudden capacity drop was linked to the occurrence of lithium deposition. The turning point of the sudden capacity drop, often referred to as "knee point" [10], was diagnosed as the state in which lithium deposition starts to become irreversible [46]. Lithium plating mechanism usually takes places in cycling conditions, and its occurrence in calendar ageing is not widely reported in the literature. According to [39], the plating phenomena in these cells could have been provoked by overcharging during the periodical characterisation tests, or uneven charge distribution within the float storage.

In some cases, a fourth phase describing a slowdown of the capacity loss was also observable (black curves since ca. 500 days in Fig. A1(b), and a green curve since ca. 300 days in Fig. A1(c), Appendix A). The references to similar observations are scarce in the literature. Petzl et al. introduced the theory of self-weakening phenomenon of the lithium plating mechanism, explaining the decrease of the ageing rate by a counter-effect of the lithium deposition [47]. Their hypothesis was that pore clogging induced by the lithium plating leads to a loss of the active material and obstructs the full charge of the cell, making electrochemically impossible for the graphite anode to reach low voltages close to the metallic lithium's voltage. This leads to a continuous reduction of the lithium plating after the turning point of the sudden capacity reduction, until the whole disappearing of the lithium plating mechanism.

In order to develop ageing models able to predict the capacity fade corresponding to Phase 3 and 4, a deep research work would be necessary to extract and validate consistent features which could explain such occurrences, as suggested in [24]. However, this requires of large amount of data, and resulted impossible with the available dataset. For this reason, modelling the Phase 3 and 4 remained out of the scope of this research work, and the corresponding data was discarded from the modelling dataset.

Therefore, in the context of this study, the modelling work focussed on capturing the relations between the storage conditions and the capacity loss of the cells, during the progressive degradation corresponding to the second phase in Fig. 1.

Besides, some unexpected trends were identified within the experimental data, for instance, a clear capacity recovery for the cells #26 and #28 at day 480 (respectively green and blue curves, Fig. A1(c), Appendix A). Such deviations are related to procedural errors during the capacity tests (e.g. exchange of the testing device, etc.). These noisy data samples could affect the performances of the model and were therefore removed from the modelling dataset.

On average, 64.64% of the initial experimental data corresponding to static ageing conditions was preserved after the preprocessing stage. The percentage of the remaining data for each cell is indicated in Table 3. Overall, all the ageing conditions of the initial experimental ageing matrix were still represented in the processed dataset. However, a large part of the cells exposed to light ageing conditions were discarded, mainly due to neglecting of the initial capacity recovery phenomenon. One of the cells stored at 35 °C and 35% SOC depicted increasing capacity values until the end of the tests and was therefore completely removed from the modelling dataset. Regarding to the cells submitted to dynamic ageing profiles, 75.56% and 52.17% of the ageing data was maintained for the cells # 31 and #32 respectively. Fig. 2 illustrates the resultant ageing data obtained after the processing stage.

## 4. Gaussian process theory

This section aims to provide a brief overview of Gaussian Process models, introducing the main concepts and the predictive equations. Detailed explanations are available in [48].

**Table 3**
Remaining data percentage ranges for each storage condition, after the data preprocessing.

| Temperature [ °C] | SOC [%] | | | | | |
|---|---|---|---|---|---|---|
| 100 | 100 | 80 | 65 | 50 | 35 | 20 |
| 25 | | 94.4 – 100% | | 14.3 – 64.7% | | |
| 35 | 66.6 – 87.5 | 95.2% | 85.7 – 90.5% | 70.0 – 90.0% | 10.0 – 50.0% | 10.0% – 15.0% |
| 45 | | 30.0 – 75.0% | | 52.9 – 88.2% | | |

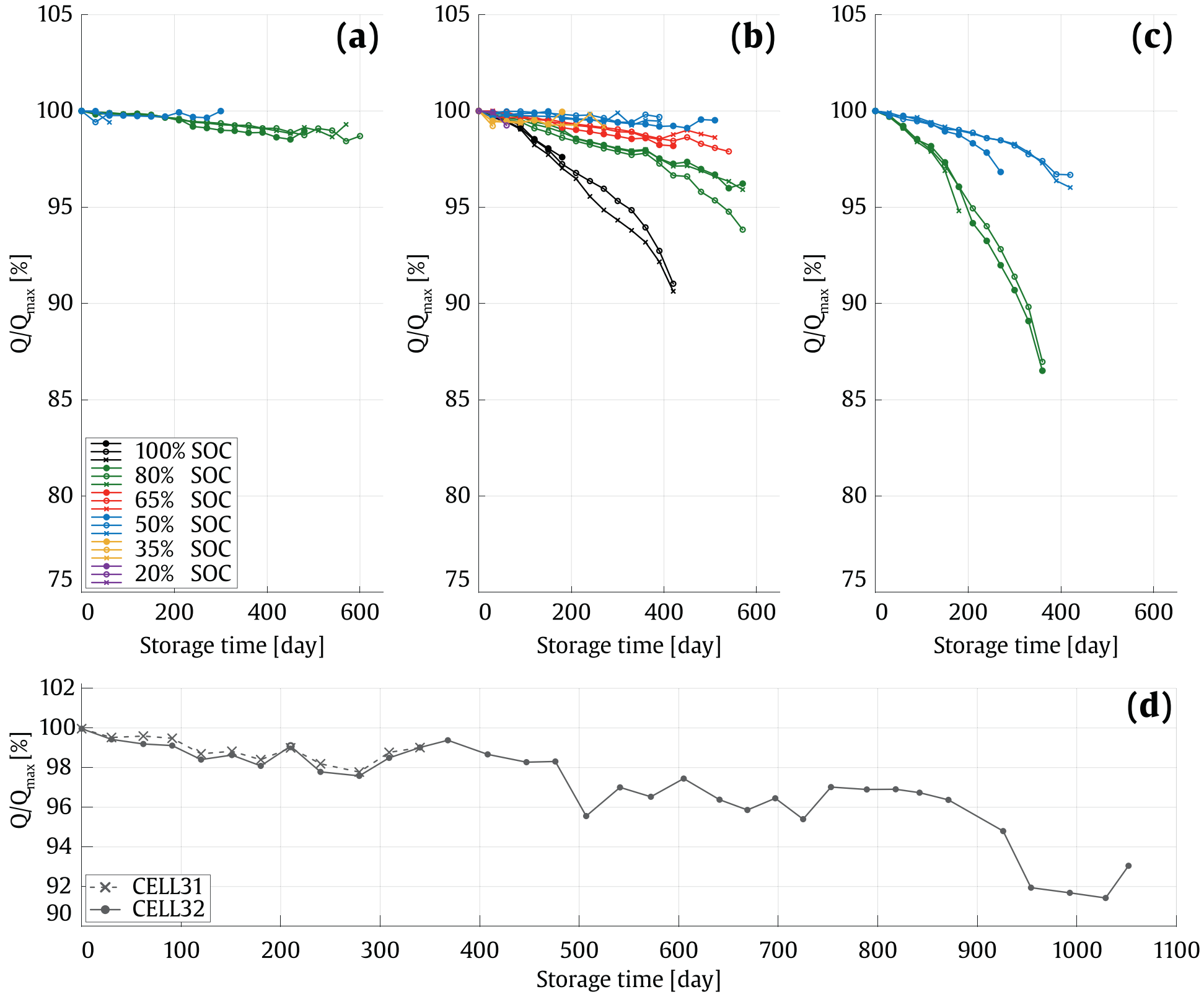

**Fig. 2.** Normalised (with maximum value $Q_{max}$) capacity, obtained after the preprocessing phase of static ageing tests at (a) 25 °C, (b) 35 °C and (c) 45 °C. (d) Normalised capacity obtained after the preprocessing phase of the dynamic ageing tests for the cells #31 and #32.

The GP is a random process, i.e. a random entity whose realisation is a function $f(\mathbf{x})$ instead of a single value. Rather than assuming a parametric form for the function to fit the data, $f(\mathbf{x})$ is assumed to be a sample of a Gaussian random process distribution. Since the GP is a nonparametric model, even when observations have been added, the model is always able to fit the new upcoming data.

A GP is fully determined by its mean and covariance functions. Defining the mean function $m(\mathbf{x})$ and the covariance function $\kappa(\mathbf{x}, \mathbf{x}')$ of a real process $f(\mathbf{x})$ as:

$$m(\mathbf{x}) = \mathbb{E}[f(\mathbf{x})]$$
$$\kappa(\mathbf{x}, \mathbf{x}') = \mathbb{E}[(f(\mathbf{x}) - m(\mathbf{x}))(f(\mathbf{x}') - m(\mathbf{x}'))] \tag{1}$$

the GP can be expressed as

$$f(\mathbf{x}) \sim \mathcal{GP}(m(\mathbf{x}), \kappa(\mathbf{x}, \mathbf{x}')) \tag{2}$$

where $\mathbf{x}$ and $\mathbf{x}'$ are two different input vectors.

Both mean and covariance functions encode the prior assumptions about the function to be learnt. They also express the expected behaviour of the model when the prediction inputs diverge from the inputs observed during training. The covariance function, also called the *kernel*, underpins the information about how relevant one target observation $y$ of the training dataset is to predict the output $y^*$, on the basis of the similarity between their respective input values $\mathbf{x}$ and $\mathbf{x}^*$.

The mean and covariance functions depend on some hyperparameters $\theta$, which must be learnt from the training dataset. From a GP point of view, the mean and covariance function selection and learning the corresponding hyperparameters are the main tasks which must be carried out during the training phase. Hyperparameters are typically estimated by the maximisation of the marginal likelihood logarithm, using the gradient of the marginal likelihood with respect to such hyperparameters [48]. The marginal likelihood is defined as the integral of the likelihood times the prior.

Under the GP framework, the prior is gaussian $\mathbf{f}|X \sim \mathcal{N}(\mathbf{0}, K)$, and the likelihood is a factorised gaussian $\mathbf{y}|\mathbf{f} \sim \mathcal{N}(\mathbf{f}, \sigma_n^2 I)$, where $\mathbf{f}$ is the vector of latent function values as $\mathbf{f} = (f(\mathbf{x}_1, \ldots, \mathbf{x}_n))^T$; $X$ is the matrix of the training input values; $\mathcal{N}$ is the gaussian (normal) distribution; $K$ is the covariance matrix for the (noise free) $\mathbf{f}$ values; $\mathbf{y}$ is the vector of the training target values; $\sigma_n^2$ is the noise variance and $I$ is the identity matrix.

The obtained log marginal likelihood is expressed in Eq. (3)

$$\log p(\mathbf{y}|X) = -\frac{1}{2}\mathbf{y}^T(K + \sigma_n^2 I)^{-1}\mathbf{y} - \frac{1}{2}\log|K + \sigma_n^2 I| - \frac{n}{2}\log 2\pi \tag{3}$$

The GP predictive equations are expressed in Eqs. (4), (5) and (6).

$$\mathbf{f}_*|X, \mathbf{y}, X_* \sim \mathcal{N}(\bar{\mathbf{f}}_*, \ \mathrm{cov}(\mathbf{f}_*)) \tag{4}$$

with

$$\bar{\mathbf{f}}_* = \mathbf{m}(X_*) + K(X_*, X)[K(X, X) + \sigma_n^2 I]^{-1}(\mathbf{y} - \mathbf{m}(X)) \tag{5}$$

$$\mathrm{cov}(\mathbf{f}_*) = K(X_*, X_*) - K(X_*, X)[K(X, X) + \sigma_n^2 I]^{-1}K(X, X_*) \tag{6}$$

where $\mathbf{f}_*$, $\bar{\mathbf{f}}_*$, and $\mathrm{cov}(\mathbf{f}_*)$ are the GP posterior prediction, its corresponding mean and its covariance, respectively; $X_*$ is the matrix of test inputs; $\mathbf{m}(X)$ and $\mathbf{m}(X_*)$ are the vectors of mean functions for the training and test inputs respectively; $K(X, X)$, $K(X_*, X_*)$, and $K(X, X_*)$ are the covariance matrices between training inputs, the test inputs, and training and test inputs, respectively.

## 5. Development of the calendar ageing model

### 5.1. Input selection

For an accurate prediction of Li-ion battery ageing at several static and dynamic operating conditions, it is necessary to consider the effect of the different stress-factors and their influence on the ageing mechanisms.

In the literature, a significant percentage of the ageing models based on Machine Learning methods do not consider the influence of such stress-factors [23]. Apart from the authors' recent work [33], only two publications were found to this effect [26,49]. Both proposed to classify and count "load patterns" depending on the stress-factors' values, defining a subset of stress-factor ranges for which the ageing is assumed to be equivalent. Then, the counted "loads" at different conditions are applied to the model. However, the definition of such "equivalency ranges" independently of the data inference could be a difficult and uncertain task, and could significantly vary from some commercial battery reference to another. The selection of too broad ranges supposes a low resolution in the input space and could lead to a poor accuracy of the model. The selection of too narrow ranges induces an increased number of inputs, increasing the computational cost and the uncertainty associated to each range.

The GP framework allows quantifying the similarities of the input space with respect to the output through data inference, depending on the kernel properties. Therefore, a better solution could be to introduce the stress-factors' values directly as an input, as already highlighted in a previous publication [33]. As a result, the "equivalency" or "similarity ranges" of each stress-factor are directly inferred from data and updated each time new data is available.

Within the context of calendar ageing, in addition to the time-dependence, the main stress-factors involved are assumed to be the cell temperature and the SOC [37]. Therefore, the model we proposed in this section considered three inputs:

- $\Delta t$: the storage time for which the ageing is predicted.
- $T^{-1}$: the reciprocal of the temperature corresponding to this storage time (for alignment to the Arrhenius law).
- *SOC*: the SOC level corresponding to this storage time.

The output of the model was the capacity loss $\Delta Q$ corresponding to a $\Delta t$ storage time at $T$ and *SOC* storage conditions.

### 5.2. Kernel construction

As explained in Section 4, the kernel $\kappa(\mathbf{x}, \mathbf{x}')$ specifies how similar or correlated the outputs $y$ and $y'$ are expected to be for two inputs $\mathbf{x}$ and $\mathbf{x}'$, respectively. The selection of the structural form of the kernel is the most important challenge in nonparametric regression [48]. However, it remains a largely subjective process based on trial and error and designer experience, as there is not any broadly accepted method to perform this task [50]. For all the GP ageing models presented in the literature, the selection of the kernel was based on trial and error methods. In this way, the kernel function presenting the lowest error with respect to a specific dataset was considered as the most suitable. Following this method, the suitability of the selected kernel in the general context of Li-ion battery ageing prediction could hardly be guaranteed, due to its high correlation to the used dataset. In order to develop GP models tailored to Li-ion battery ageing application, a stronger justification of the kernel selection is desirable.

As noted in Section 5.1, the model must be able to handle different input dimensions. Consequently, compositional kernels' framework is a suitable solution to construct a main kernel composed of interpretable components, each one related to a specific input dimension [50]. In order to focus on the behaviour of the composed kernels, a zero-mean function was defined in this work. This is not a significant limitation, since the mean of the posterior process is not confined to be zero [48].

#### 5.2.1. Selecting individual kernel components

As explained in Section 4, the GP framework is a nonparametric model, and therefore the *learning* problem is the problem of finding the suitable properties of the function (isotropy, anisotropy, smoothness, etc.), rather than a particular functional form [48].

The range of the SOC input dimension is intrinsically limited between 0 and 100%. This is defined to be a local modelling problem. In the context of the development of ageing models oriented to learn from the data observed after their deployment in real application, the definition of the similarity using the Euclidean distance seems suitable, as it could allow the model to cover the whole SOC range after the observation of a few data points. Therefore, the kernel components corresponding to the SOC input space could be represented by *isotropic* kernels. Furthermore, the operation window corresponding to the temperature input is limited by the recommendations of the manufacturer (i.e. storage temperatures between −30 °C and 55 °C), specified in Table 1. Accordingly, isotropic kernel could also be assigned to such input dimensions. Furthermore, different kind of isotropic kernels could be selected for these inputs, depending on the smoothness assumption for the process. The Ornstein-Uhlenbeck kernel, detailed in [48], was deemed too rough to describe the influence of the stress-factors on ageing. Besides, although the squared-exponential kernel is the most widely used isotropic kernel, its strong smoothness assumption was claimed to be unrealistic for modelling many physical processes (e.g. implication of charging C-rate deviations on underlying capacity loss) and the Matérn kernel class was recommended instead [48]. Therefore, a 5/2 Matérn kernels were selected to host independently the input dimensions corresponding to each stress-factor.

The kernel component related to the Δt input dimension requires several Δt values to be involved in the training data, in order to optimise the associated hyperparameters. In order to limit the training computation time, only three different values of Δt were processed in the training data (which are 30, 60 and 90 days). Table 4 illustrates the structure of the training data. In this context, the use of an isotropic kernel requires a large amount of different values of Δt for long-term prediction, implying a large quantity of training data and increased computation times. Therefore, this kernel component should be anisotropic. In the second phase of the Li-ion cells ageing described in Fig. 1, the capacity loss seems to be linear with respect to Δt. Therefore, a linear kernel component was selected for this input dimension.

Although the data vectors "CELL002 – data vector 1" and "CELL002 – data vector 4" in Table 4 have the same inputs values, the target is different because both correspond to a the capacity loss from a different starting point, in the capacity curve of the CELL002. The data vectors with identical input values and different outputs are useful for the determination of the noise hyperparameter of the GP models (see Eq. (7)).

#### 5.2.2. Composing the whole kernel

In the GP framework, the kernel function is also a covariance function and therefore must be positive semidefinite [48]. Moreover, positive semidefinite compositional kernels are closed under the addition and multiplication of basic kernels. The effect of these operations is well explained in [50], for example: *"A sum of kernels can be understood as a* [logical] *OR operation. Two points are considered similar if either*

**Table 4**
Example of the training data structure.

| | | Input vector $\mathbf{x}$ | | | Target $y$ |
|---|---|---|---|---|---|
| | | $\Delta t$ [days] | $T^{-1}$ $[K^{-1}]$ | $SOC$ [%] | $\Delta Q$ [%] |
| CELL02 | data vector 1 | 30 | 0.0034 | 80 | −0.041 |
| | data vector 2 | 60 | | | −0.136 |
| | data vector 3 | 90 | | | −0.181 |
| | data vector 4 | 30 | | | −0.095 |
| | data vector 5 | 60 | | | −0.140 |
| … | … | … | … | … | … |
| CELL09 | data vector 1 | 30 | 0.0032 | 100 | −0.310 |
| | data vector 2 | 60 | | | −0.572 |
| | data vector 3 | 90 | | | −0.949 |
| | data vector 4 | 30 | | | −0.261 |
| | data vector 5 | 60 | | | −0.638 |
| … | … | … | … | … | … |

*kernel has a high value. Similarly, multiplying kernels is a* [logical] *AND operation, since two points are considered similar only if both kernels have high values".*

Additive kernels assume the added stochastic processes to be independent. However, the inputs $T$ and $SOC$ interact in the kinetics reactions inside the electrode [15], hence additive kernel composition should be avoided. In order to account for the interactions between the different input dimensions, the tensor product is suggested within [48,50] and is used in the composed kernel (Eq. (7)).

$$\kappa(\mathbf{x},\mathbf{x}') = \sigma_f{}^2 \cdot \left[ \begin{array}{c} \left(1+\sqrt{5}\cdot\frac{|x_1-x'_1|}{\theta_T}+\frac{5}{3}\cdot\frac{|x_1-x'_1|^2}{\theta_T{}^2}\right)\cdot\exp\left(-\sqrt{5}\cdot\frac{|x_1-x'_1|}{\theta_T}\right) \\ \cdot\left(1+\sqrt{5}\cdot\frac{|x_2-x'_2|}{\theta_{SOC}}+\frac{5}{3}\cdot\frac{|x_2-x'_2|^2}{\theta_{SOC}{}^2}\right)\cdot\exp\left(-\sqrt{5}\cdot\frac{|x_2-x'_2|}{\theta_{SOC}}\right) \\ \cdot\,(x_3\cdot x'_3 + \theta_t{}^2) \end{array} \right] + \sigma_n{}^2\cdot I \tag{7}$$

where $\mathbf{x}$ and $\mathbf{x}'$ are different input vectors structured as $\mathbf{x} = (x_1,\ x_2,\ x_3)$, with $x_1 = T^{-1}$, $x_2 = SOC$, $x_3 = \Delta t$; $\theta_T$, $\theta_{SOC}$, and $\theta_{\Delta t}$ are the hyperparameters related to the $T$, $SOC$ and $\Delta t$ inputs respectively. The additional hyperparameters $\sigma_f^2$ and $\sigma_n{}^2$ are respectively the signal variance, which plays the role of scaling the outputs in the dimension of the capacity loss $\Delta Q$, and the noise variance, which models an additive Gaussian noise from the data.

## 6. Learning from static operating conditions

This section aims to illustrate the ability of the developed GP model to improve its prediction performances while observing an increasing number of battery calendar operation data. Indeed, as new observations of storage conditions are presented to the model, the training dataset of the model involves a more comprehensive view of the influence of the different combinations of stress-factors on the capacity loss. Therefore, for each prediction, the covariance function is able to find more similar examples in the training dataset, in term of storage conditions. The prediction performances of the model improve throughout the whole operation window of the Li-ion cells.

In this section, the improvement of the model performances was evaluated in terms of:

(i) *Accuracy of the prediction*: as the training dataset increases, a reduction of the prediction errors is expected over the whole operation window. The metrics used to evaluate the prediction error are detailed in Section 6.1.
(ii) *Confidence in the prediction*: as the training dataset increases, the model disposes of more information about the ageing throughout the whole operation window. In accordance with the covariance equation Eq. (6), the confidence intervals of the predictions are expected to reduce, signifying that the model is more confident about its predictions. The metric used to evaluate the accuracy of the confidence intervals is detailed in Section 6.1.

### 6.1. Evaluation metrics

Six different metrics were used to assess the prediction performances of the two ageing models. The first one was the root-mean-square error (RMSE) of the output of the model, which was the capacity loss $\Delta Q$, defined according to Eq. (8).

$$RMSE_{\Delta Q}(\hat{y}_i, y_i) = \sqrt{\frac{1}{N_T}\sum_{i=1}^{N_T}(\hat{y}_i - y_i)^2} \tag{8}$$

where $\hat{y}_i$ is the predicted output, $y_i$ is the measured output and $N_T$ is the number of points to be evaluated. The second metric was defined as the RMSE of the predicted capacity curve:

$$RMSE_Q(\hat{Q}_i, Q_i) = \sqrt{\frac{1}{N_T}\sum_{i=1}^{N_T}(\hat{Q}_i - Q_i)^2} \tag{9}$$

where $\hat{Q}_i$ is the predicted capacity calculated by accumulation of the output and $Q_i$ is the measured capacity. This second metric is useful in order to evaluate the accumulative error of the model.

The RMSE is useful to assess the prediction performances of a model, with an emphasis on the high deviations which are strongly penalised. In order to evaluate the ability of the model to capture the main trends of the data, the analysis was completed with the implementation of the mean-absolute-error (MAE), defined in Eqs. (10) and (11) in terms of model output and capacity curve, respectively.

$$MAE_{\Delta Q}(\hat{y}_i, y_i) = \frac{1}{N_T}\sum_{i=1}^{N_T}|\hat{y}_i - y_i| \tag{10}$$

$$MAE_Q(\hat{Q}_i, Q_i) = \frac{1}{N_T}\sum_{i=1}^{N_T}|\hat{Q}_i - Q_i| \tag{11}$$

In the context of this study, the main objective of the model was to capture the main trends of Li-ion battery ageing in different operating conditions, rather than achieving a perfect fit of each data point. Therefore, a 2% $MAE_Q$ threshold was defined as acceptable prediction error.

The final metric was the calibration score, which aimed at quantifying the accuracy of the uncertainty estimates. It is defined as the percentage of measured results in the test dataset that are within a predicted credible interval. Within a $\pm\ 2\sigma$ interval, corresponding to a 95.4% probability for a Gaussian distribution, the calibration score is given by Eqs. (12) and (13).

$$CS_{2\sigma-\Delta Q} = \frac{1}{N_T}\sum_{i=1}^{N_T}[|\hat{y}_i - y_i| < 2\sigma]\cdot 100 \tag{12}$$

$$CS_{2\sigma-Q} = \frac{1}{N_T}\sum_{i=1}^{N_T}[|\hat{Q}_i - Q_i| < 2\sigma]\cdot 100 \tag{13}$$

Therefore, $CS_{2\sigma}$ should be approximately 95.4% if the uncertainty predictions are accurate. Higher or lower scores indicate under- or over-confidence, respectively [26].

### 6.2. Training case studies to illustrate the learning of new operating conditions

In order to illustrate how the GP model could learn from new observations and improve prediction performances, 7 distinct training cases were defined, each one involving a different number of training data from the ageing dataset presented in Section 3. From the training case 1 to the training case 7, the number of training data increased: the data corresponding to new storage conditions was included progressively, revealing one by one the influence of the different levels of the different stress-factors.

Accordingly, the distinct temperature values were introduced from case 1 to case 2, followed by the different SOC levels from case 3 to case 7. The training case 1 involved the single 80% SOC condition at the temperature extrema of the static test matrix (25 °C and 45 °C). The temperature range was completed in case 2 with the additional value of 35 °C. Since the training case 3, different SOC storage values were introduced, starting by the 50% SOC value at the three temperatures. The SOC range was then progressively completed alternating the incorporation of highest and lowest values, i.e. 100%, 20%, 65% and 35% SOC respectively in training cases 4, 5, 6, and 7. The characteristics of each training case are summarised in Table 5, specifying the different storage conditions involved during the training process, as well as the corresponding ratio of the amount of training data with respect to the whole available data.

**Table 5**
Summary of the different case studies, specifying the different cells involved and the related storage conditions, as well as the ratio of the amount of training data with respect to the whole available data.

| | | Learning Temperature | | | | Learning SOC | | | | | | | # Training data / # Total data [%] |
|---|---|---|---|---|---|---|---|---|---|---|---|---|---|
| **CASE 1** | **T** | 25 | | 45 | | | | | | | | | 24.86 |
| | **SOC** | | 80 | | | | | | | | | | |
| **CASE 2** | **T** | 25 | | 45 | 35 | | | | | | | | 42.40 |
| | **SOC** | | | 80 | | | | | | | | | |
| **CASE 3** | **T** | 25 | | 45 | 35 | 25, 35, 45 | | | | | | | 70.13 |
| | **SOC** | | | 80 | | 50 | | | | | | | |
| **CASE 4** | **T** | 25 | | 45 | 35 | 25, 35, 45 | 35 | | | | | | 79.70 |
| | **SOC** | | | 80 | | 50 | 100 | | | | | | |
| **CASE 5** | **T** | 25 | | 45 | 35 | 25, 35, 45 | | 35 | | | | | 80.44 |
| | **SOC** | | | 80 | | 50 | 100 | | 20 | | | | |
| **CASE 6** | **T** | 25 | | 45 | 35 | 25, 35, 45 | | | 35 | | | | 95.43 |
| | **SOC** | | | 80 | | 50 | 100 | | 20 | | 65 | | |
| **CASE 7** | **T** | 25 | | 45 | 35 | 25, 35, 45 | | | | 35 | | | 100 |
| | **SOC** | | | 80 | | 50 | 100 | | 20 | | 65 | 35 | |

### 6.3. Prediction results

#### 6.3.1. Accuracy improvement

The black curves in Fig. 3 indicate the prediction accuracy of the GP model proposed in Section 5, trained with the different training cases defined in Section 6.2, in term of $MAE_{\Delta Q}$ and $MAE_Q$. The corresponding RMSE values are indicated in Table B1, Appendix B. For each training case, the error calculation was performed separately for:

(i) The training cells: the mean value of the prediction errors obtained for all the cells involved in the training case was calculated (Fig. 3(a)). Such errors are informative about the ability of the model to fit the training data.
(ii) The validation cells: the mean value of the prediction errors obtained for all the cells not involved in the training case was calculated (Fig. 3(b)). Such error is relevant to evaluate the generalisation ability of the model.
(iii) All the cells: the mean value of the prediction errors obtained for all the cells (Fig. 3(c)). Such error is informative about the global accuracy of the model.

As expected, the predictions errors of the training cells in Fig. 3(a) fulfil the 2% $MAE_Q$ threshold for all the training cases. Regarding the validation cells, the threshold of the 2% $MAE_Q$ is reached for the training case 2 (see Fig. 3(b)). For the training case 3, the model achieved 0.64% $MAE_Q$ accuracy and the performances of the model seem not to improve significantly since such training case.

Fig. 4(a–d) illustrates the capacity predictions of the GP model resulting from the training case 3, for different storage conditions involved in the training data. The average $MAE_{\Delta Q}$ and $MAE_Q$ errors of the model corresponding to the training case 3 were 0.27% and 0.47%, respectively, for the training cells. The average $CS_{2\sigma-\Delta Q}$ and $CS_{2\sigma-Q}$

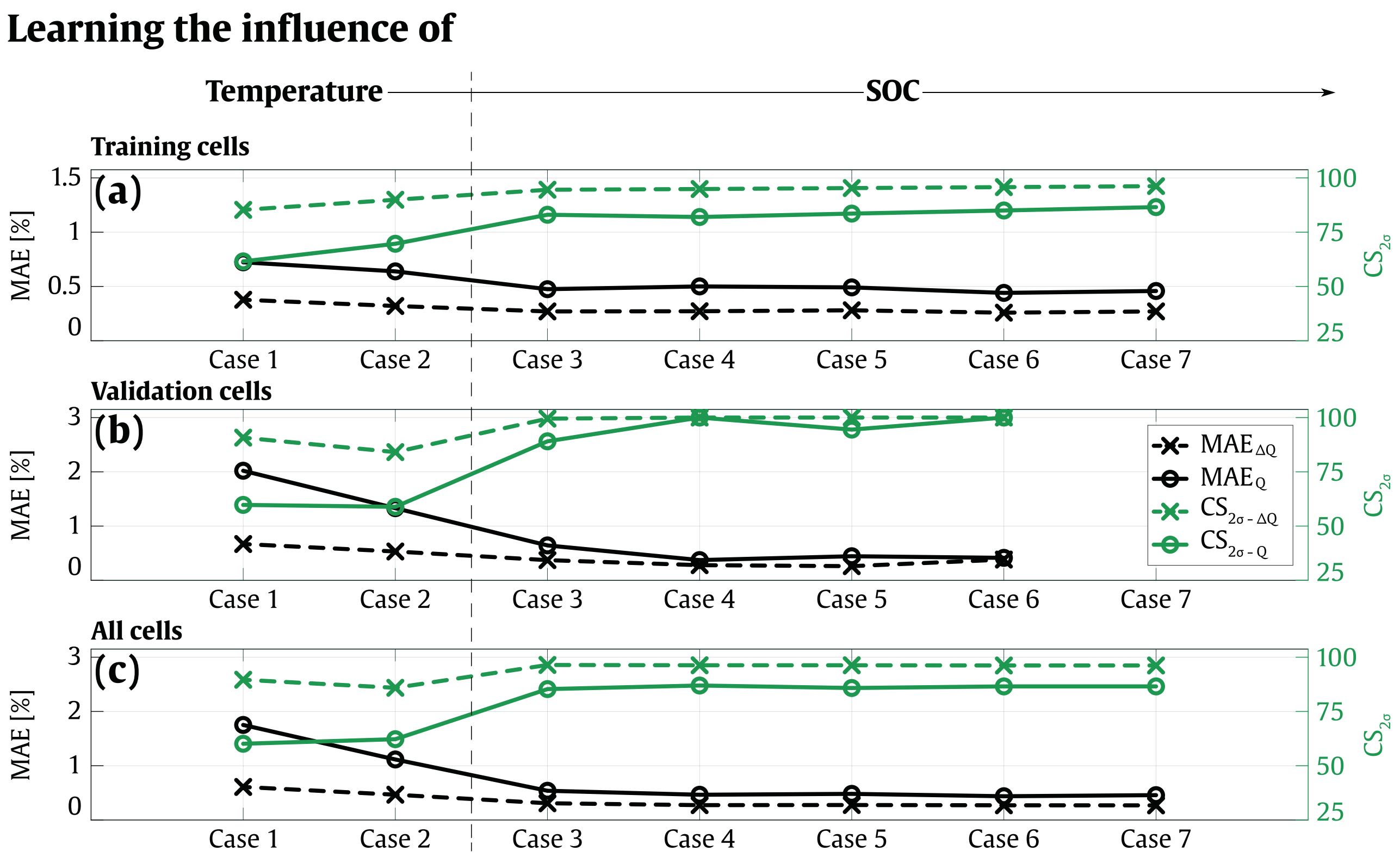

**Fig. 3.** Prediction results corresponding to each training case, in term of $MAE_Q$ and $CS_{2\sigma}$, distinguishing the errors of (a) all the training cells, (b) all the validation cells and (c) all the cells.

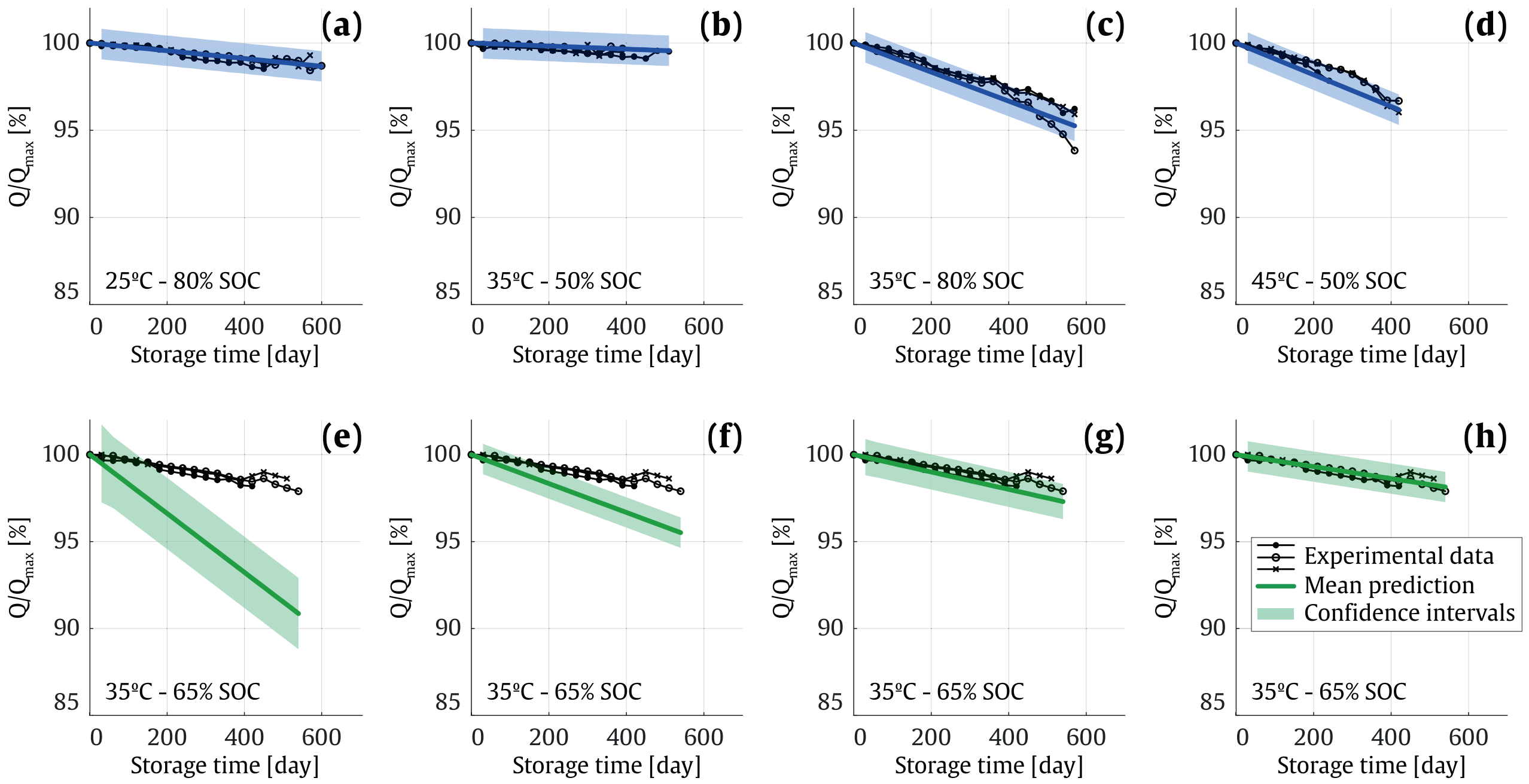

**Fig. 4.** Capacity predictions with the GP model trained at training case 3, for the training cells stored at (a) 25 °C and 80% SOC, (b) 35 °C and 50% SOC, (c) 35 °C and 80% SOC, (d) 45 °C and 50% SOC. Capacity predictions for the cells stored at 35 °C and 65% SOC, with the GP models trained at (e) training case 1, (f) training case 2, (g) training case 3 and (h) training case 7.

were respectively 94.54% and 83.03%.

Fig. 4(e–h) aims to underpin the improvement of the generalisation performances of the GP, while increasing the number of training values in the input space of the SOC. To this end, the capacity predictions were represented for the cells stored at 35 °C and 65% SOC, using GP models obtained from different training cases. The model obtained from the training case 1 did not have any information neither about the degradation at 35 °C nor about the effect of SOC on the capacity loss, as the training data involved the single input of 80% SOC. At this stage, the prediction at lower SOC levels were over-estimated (see Fig. 4.(e)). The mean error obtained at 35 °C and 65% SOC storage condition was 3.15% $MAE_Q$. In the training case 2, the incorporation of the 35 °C storage temperature in the training data improved significantly the prediction, reaching a 1.12% $MAE_Q$, Fig. 4.(f). In the training case 3, the model started to learn the effect of the SOC by incorporating a 50% SOC condition in the training data. The mean error of the prediction improved drastically (0.34% $MAE_Q$), as the model could infer from two different SOC values and gain a numerical intuition about the effect of the SOC on capacity loss (see Fig. 4(g)). For comparison, the results obtained with a fully trained GP (training case 7) were also plotted in Fig. 4.(h): there was not significant improvement in term of error reduction. However, the confidence intervals were slightly reduced, indicating a higher confidence of the model to perform predictions in at 65% SOC, since such operating condition was represented in the training data (more details in Section 6.3.2). At this point, it is noteworthy that the model corresponding to the training case 7 is only used in this study for a sake of comparison with the previous cases. In fact, such a model would be unreliable for deployment, as it involves all the available data for training and then its performances would not be further validated under static conditions.

### 6.3.2. Increase of confidence

As stated by the variance equation Eq. (6), the confidence intervals of a prediction reduce if the training dataset involves data samples similar to the predicted input values. Informally, this means that the model feels more confident to do predictions in case it already observed similar operating conditions in training data. Therefore, the analysis of the width of the confidence intervals – or equivalently the standard deviation value - along a large operating range of each stress-factor is informative about how confident the model feels to perform predictions throughout a broad operating window. In this sense, the evolution of the standard deviation throughout the input space testifies about the learning process of the model.

Fig. 5 shows the evolution of the standard deviation of the GP model predictions throughout the whole operation window of the Li-ion cell under study, for the different training cases. In Fig. 5(a), the standard deviation of the model obtained from the training case 1 indicates lowest values around 25 °C and 45 °C, which are the only storage temperatures experienced at this stage. The observation of the effect of a 35 °C operation in the training case 2 flattened the curve around such temperature: at this stage, the obtained model felt relatively confident to perform predictions within the 20 °C - 50 °C temperature range. It is noteworthy that the model presented high standard deviation values at low and negative temperatures, due to the lack of information in such storage regions. Fig. 5(b)–(e) corresponds to the learning of the influence of the SOC, showing the evolution of the standard deviation of the GP model predictions throughout the whole SOC range and at constant 15 °C, 25 °C, 35 °C and 45 °C, respectively. As expected, the lowest standard deviation stood near 50% and 80% for training case 3, and the observation of intermediate SOC levels from the training cases 4 to 7 lead to reduced values in the whole range, unless below 20% SOC operation which still was an unknown storage condition. It is noteworthy that the lowest standard deviation values are observable at 35 °C, Fig. 5(d), as the SOC input space was explored at this temperature. Besides, the standard deviation of the predictions at 15 °C, Fig. 5(b), achieved highest values for the training case 7. This is due to the higher relevance associated to the temperature in such training case (see sensitivity analysis in Section 6.3.3), which led to a higher gradient in the evolution of the standard deviation throughout unknown temperatures, e.g. colder temperatures than those involved in the training dataset.

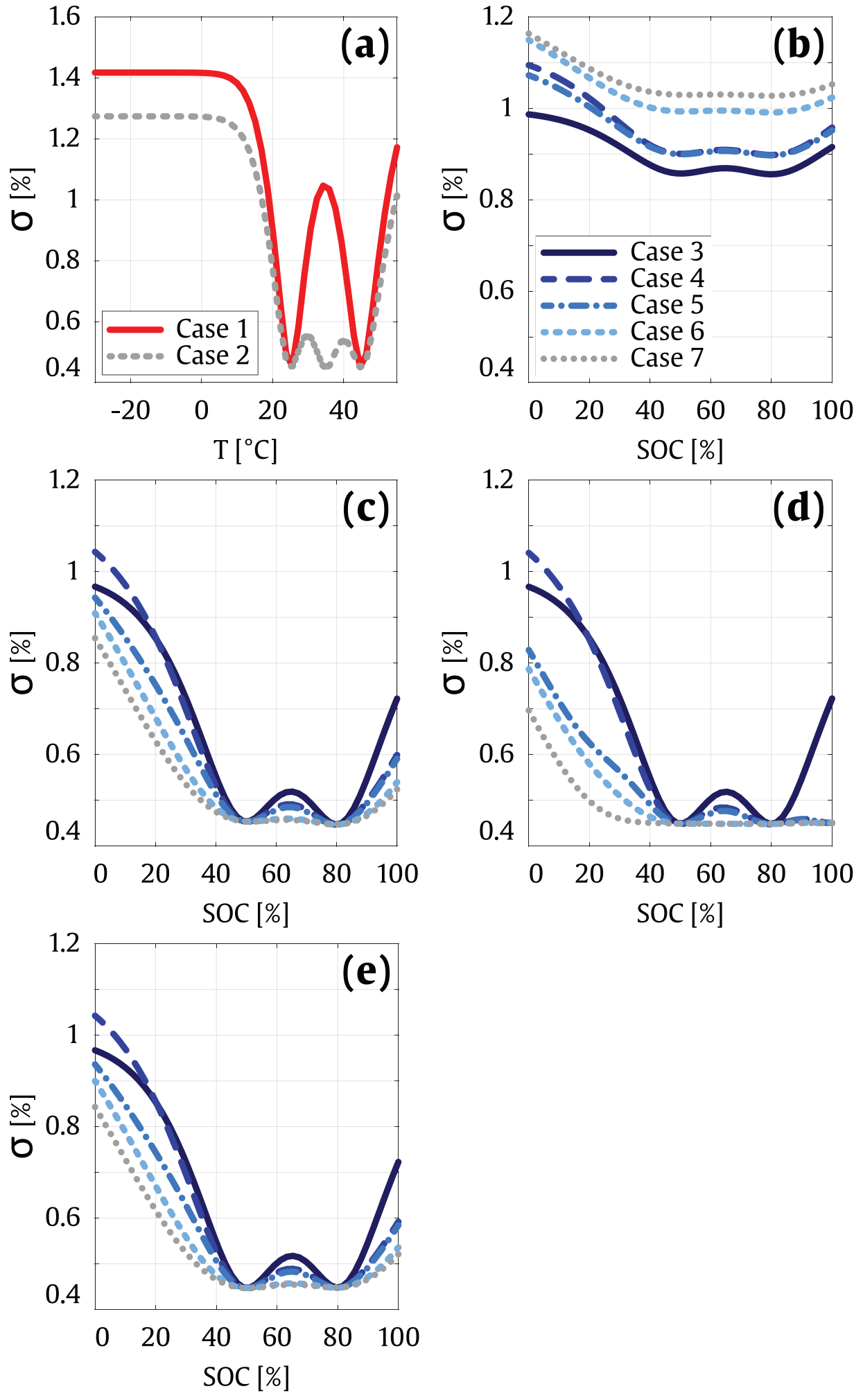

**Fig. 5.** Evolution of the standard deviations of the GP model predictions throughout the whole operation window of the Li-ion cell under study, from training case 1 to 7. (a) Evolution throughout the temperature space, at constant 80% SOC (b) Evolution throughout the SOC space, at constant 15 °C, (c) Evolution throughout the SOC space, at constant 25 °C, (d) Evolution throughout the SOC space, at constant 35 °C and (e) Evolution throughout the SOC space, at constant 45 °C.

The reduction of the standard deviation in Fig. 5 testifies about the increment of the model's confidence to perform prediction throughout a broad operating window, as input spaces are progressively explored. Furthermore, the accuracy of the confidence level of the model was evaluated using the calibration score metric, introduce in Section 6.1. As previously explained, the $CS_{2\sigma}$ values should be approximately 95.4% if the uncertainty predictions are accurate. Higher or lower scores indicate under- or over-confidence of the model, respectively [26].

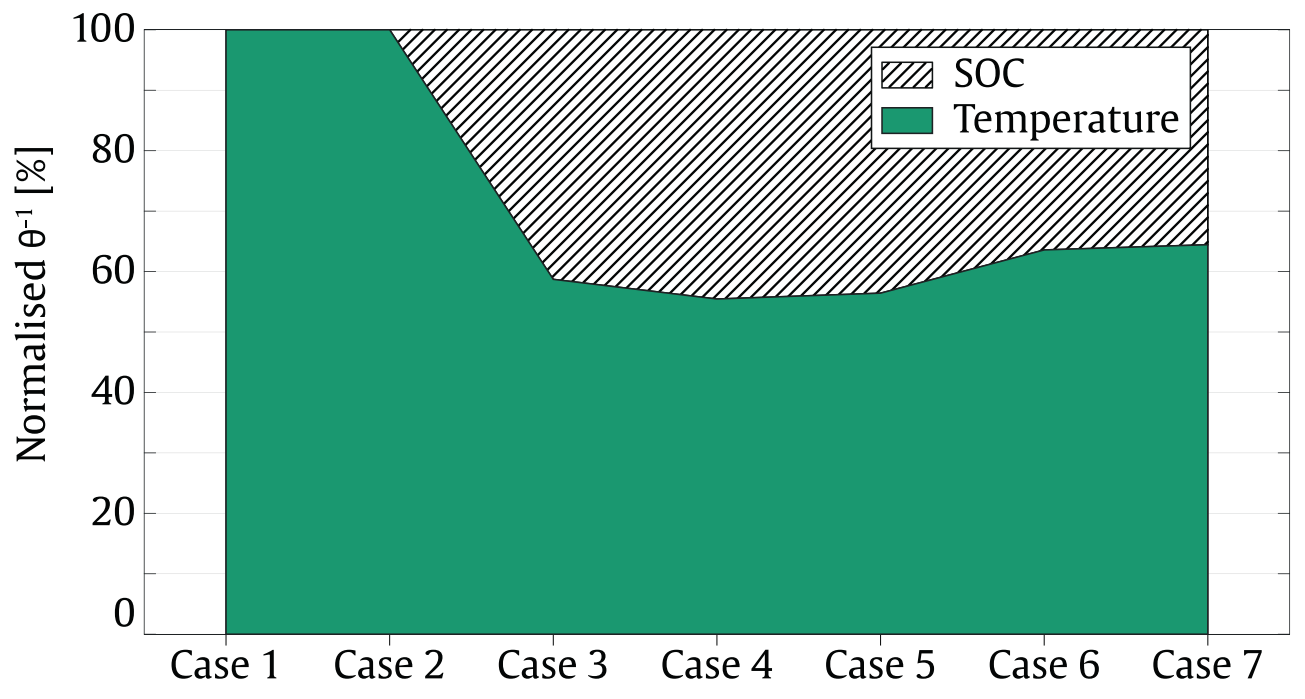

**Fig. 6.** Evolution of the relative relevance of the different stress-factors, from the training case 1 to 7.

In Fig. 3, the evolution of the mean value of the calibration scores were plotted for each training case of the GP model, in term of capacity loss and accumulated capacity. Since the training case 3, the overall $CS_{2\sigma-Q}$ values converge into approximately 86.55% (Fig. 3(c)). This traduces a slightly over-confident behaviour of the model in term of the accumulated capacity. However, regarding the calibration scores values corresponding to the output of the model, the overall $CS_{2\sigma-\Delta Q}$ values converge into approximately 96.19%.

### 6.3.3. *Sensitivity of the capacity loss to the stress-factors*

For many covariance functions, the observation of the hyperparameters allows one to interpret how the GP model understand the data. For isotropic kernels, the hyperparameters play the role of characteristic length-scale. Such covariance functions implement automatic relevance determination, since the inverse of the length-scale determines how relevant an input is: if the length-scale has a very large value, the covariance will become almost independent of that input, effectively removing it from the inference [8]. Therefore, the sensitivity of the capacity loss to the different stress-factors could be analysed by observing the inverse of their respective hyperparameters. Fig. 6 displays, for each training case, the inverse of the hyperparameters corresponding to the temperature and SOC, relatively normalised to each other.

In the training cases 1 and 2, only the temperature involved different storage values in the training dataset, as the single value of 80% was available for the SOC input. In absence of data to guide the optimisation of the corresponding hyperparameters, a high initial hyperparameter value was imposed to the SOC input, in order to hinder its optimisation and then remove its effect from inference. In this context, the unique relevant stress-factor for the GP model was the temperature.

From the training case 3 to 7, different SOC levels were progressively included in the training dataset, and the corresponding hyperparameter was 'released' for optimisation. In Fig. 6, it could be observed that the relative relevance of the SOC input with respect to the capacity loss increased for the training case 3; however, the temperature variations was still considered slightly more impactful on the capacity loss than SOC variations. The training case 4 included the data corresponding to 100% SOC storage condition, which present a relatively high acceleration of the capacity loss: this increased the relative weight of the SOC with respect to the temperature. The following cases 5, 6 and 7 included the data corresponding to the lower SOC levels, highlighting that the variation of SOC at its low values has a reduced effect on capacity loss (observable in Fig. 2(b)). This resulted in the mitigation of the relative relevance of the SOC input, compared to the temperature. At this point, it is important to highlight that although such comparison could clarify how the GP model understand the data, it does not imply causality.

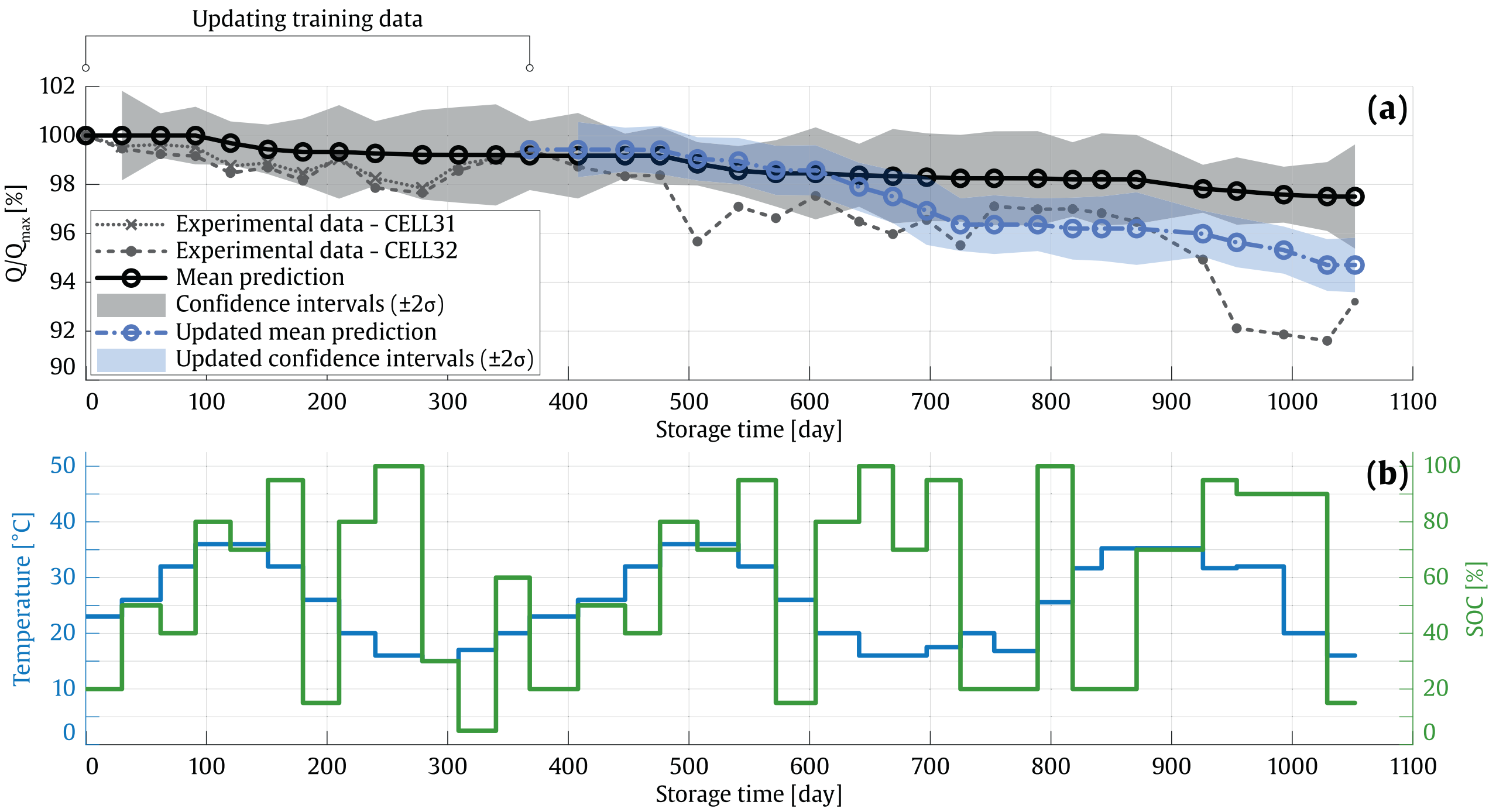

**Fig. 7.** (a) Normalised capacity (with initial value $Q_{max}$) of the cells #31 and #32, after the preprocessing of the raw data obtained from the dynamic ageing tests (dotted grey lines) and the corresponding ageing predictions for the initial model (training case 3, black line and grey area) and the updated model (blue line and area). (b) Storage temperature and SOC dynamic profiles, applied during the dynamic ageing tests for the cells #31 and #32.

## 7. Learning from dynamic operating conditions

The operating conditions of Li-ion batteries are barely constant in real applications. This implies that the ageing models developed in the basis of ageing tests realised at constant operating conditions must be validated at dynamic operating conditions. Furthermore, as this study focusses on the development of ageing models oriented to learn from ageing data collected from real-world operation, the analysis of the possibility to infer about the correlations among the different stress-factors and the capacity loss directly from dynamic operation profile is necessary.

For this purpose, the model developed in Section 5 was employed to perform ageing predictions for cells #31 and #32, the operating profiles of which were presented in Fig. 7(b). In Section 6, the GP model reached satisfying prediction results for the training case 3 achieving an overall error of 0.53% $MAE_Q$, and the performances of the model did not improve significantly since such training case. In this section, such training case was therefore selected as initial state of the model, in order to evaluate the prediction performances of the model at dynamic operating conditions. The obtained predictions are presented in black line (mean prediction) and grey area (confidence intervals) in Fig. 7(a), for the cells #31 and #32.

The errors of the predictions for the model obtained from training case 3 were 0.72% and 0.42% in terms of $MAE_{\Delta Q}$, and 1.78% and 0.62% in terms of $MAE_Q$, for the cells #31 and #32 respectively. At approximately 368 days in storage, the whole range of the temperature profile was experienced for the cells #31 and #32, Fig. 7(b). For the cell #32, different combinations of the temperature and SOC level were also observed, some of them reproduced on the remaining storage profiles (e.g. the combinations between ca. 0 and 368 days were reproduced between ca. 368 and 641 days). Such point was then deemed to be a suitable updating point for the model, to be able to evaluate the learning ability of the model at dynamic operating conditions. Therefore, the operating conditions as well as the corresponding capacity loss values observed between 0 and 368 days were included in the training dataset in order to obtain an updated GP model.

The predictions performed with the updated model were represented in blue in Fig. 7(a), for cell #32. The initial model predicted larger confidence intervals at cold temperatures (between 15 °C and 25 °C), as the coldest temperature experienced in the training case 3 was 25 °C. The inclusion of such values in the training set increased the confidence of the model to perform predictions in this range. This is traduced in Fig. 7(a) by reduced confidence intervals at cold temperatures, compared with the initial predictions.

When updating the model with the different temperature and SOC combinations observed in the dynamic profiles, the confidence of the model for predicting throughout the whole window of the storage conditions accordingly improved. This is observable in Fig. 8, which reflects the evolution of the standard deviation of the model's predictions, for the model corresponding to the training case 3 and the model updated with the data obtained from dynamic storage profiles until 368 days. Regarding the range of the storage temperatures, Fig. 8(a), it is remarkable that the model gained confidence at coldest temperatures, which is reflected by a reduction of the standard deviation in such region. This is also observable in Fig. 8(b), which indicates the standard deviations of the predictions at constant 15 °C and throughout the whole SOC range: the models' predictions are clearly more confident at such temperatures, with minimal values around 35% and 100% SOC levels, which are the values at which the cell was stored when experiencing cold temperatures (Fig. 7(b), between 240 – 309 storage days). Fig. 8(c), (d) and (e) depicts the evolution of the standard deviation respectively at 25 °C, 35 °C and 45 °C. It could be observed that the model gained confidence notably between the range of 0% - 40% SOC storage conditions.

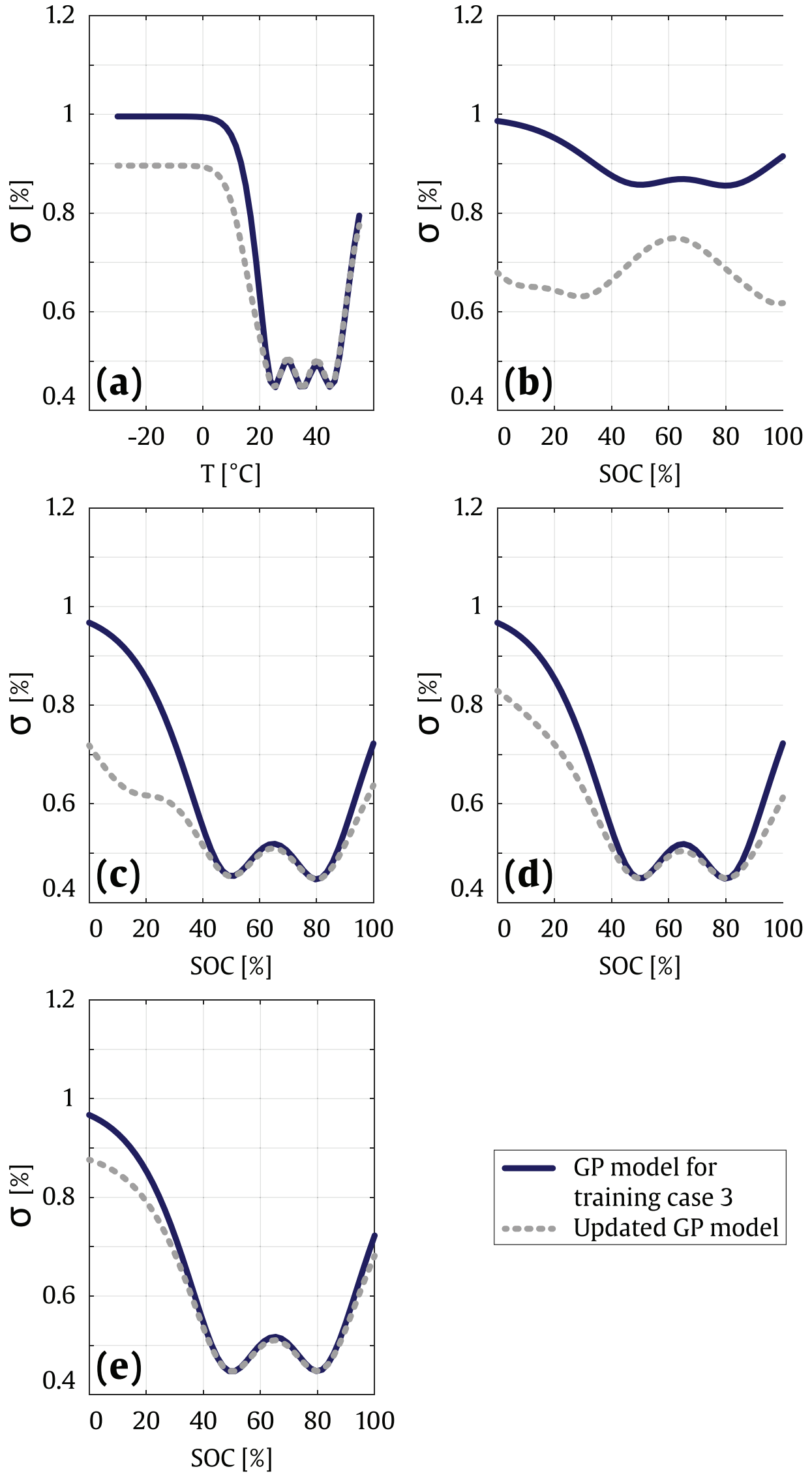

**Fig. 8.** Evolution of the standard deviations of the GP model predictions throughout the whole operation window of the Li-ion cell under study, from the model trained at case 3 and the model updated at dynamic operating conditions. (a) Evolution throughout the temperature space, at constant 80% SOC (b) evolution throughout the SOC space, at constant 15 °C, (c) evolution throughout the SOC space, at constant 25 °C, (d) evolution throughout the SOC space, at constant 35 °C and (e) evolution throughout the SOC space, at constant 45 °C.

## 8. Discussion, limitation of the study and further works

The model developed in Section 5 demonstrated suitable performances to fit the data, independently from the number of training data and involved stress-factors. This is observable in Fig. 3(a), where both $MAE_{\Delta Q}$ and $MAE_{Q}$ curves of the training cells showed a constant level under the defined 2% threshold, from the training case 1 to 7.

Regarding the amount of experimental ageing tests necessary from the laboratory for the development of the initial ageing model, the training case 3 seems to present an adequate trade-off between the performances and the development cost of the model, insofar as the cell is used at the operating conditions recommended by the manufacturer (Table 1). In fact, the model achieved an overall error of 0.53% $MAE_{Q}$, which is below the defined 2% $MAE_{Q}$ threshold, and the performances of the model seem not to improve significantly since such training case (see Fig. 3(b) and (c)).

The reduction of the standard deviation in Figs. 5 and 8 testified about the increment of the model's confidence to perform prediction throughout a broad window of the storage conditions, as the temperature and SOC input spaces are progressively explored. However, the developed GP model turned out to be slightly over-confident, according to the calibration scores curves represented in Fig. 3. As previously explained, the $CS_{2\sigma}$ values should be approximately 95.4% if the uncertainty predictions are accurate: the obtained $CS_{2\sigma-Q}$ and $CS_{2\sigma-\Delta Q}$ values converged approximately into 86.55% and 96.19% respectively (Fig. 3(c)). It could be observed that the confidence intervals of the model output are relatively close to the target value of 95.4%. The difference between the $CS_{2\sigma-Q}$ and $CS_{2\sigma-\Delta yQ}$ suggests that the over-confidence of the model is induced by the error accumulation of the iterative prediction process. Therefore, further investigations would be required in order to study the propagation of model's uncertainty throughout the long-term ageing prediction [51].

In Section 7, the developed model was validated at dynamic operating conditions, and the ability of the model to learn directly from dynamic operating conditions was illustrated with 2 cells. This study should be extended involving more cells stored at dynamic conditions. Additionally, the validation procedure should be completed by verifying the performance of the models at realistic storage conditions, involving daily temperature fluctuations [52]. The author's plan to address this important topic in further research, based on extensive experimental ageing data obtained for different real operating scenarios.

Before the training of the developed model, the data was submitted to a preprocessing stage (Section 3). It was assumed that the initial capacity rise (Phase 1) identified in the capacity curve of several cells is not provoked by any ageing mechanism and does not have any influence on the further ageing trend of the cells. These assumptions should be verified in further work. Furthermore, the preprocessing method presented in Section 3 requires a slightest knowledge of the ageing trajectory, in order to distinguish the different phases in the capacity curves. Within the context of model deployment, the capacity data are obtained one by one according to the progressive ageing of the cells, and it could be difficult to classify each point immediately within the different phases. However, it is noteworthy that the model updating procedure should not necessarily be performed immediately after the obtention of each new data point, as the ageing of a Li-ion cell is a relatively slow procedure (especially on an early degradation stage). In fact, a "mini-batch" training approach consisting on waiting until the observation of several data points could be a better strategy: this could allow to assess the relevance of the upcoming data points, minimise data acquisition deviation, and classify correctly the corresponding capacity curve phases, which is necessary to the appropriate modelling of the target degradation mechanisms.

Moreover, cycle-induced calendar ageing is another important source of capacity degradation. If the cells are severely degraded due to cycling, they may suffer more capacity degradation even under moderate storage conditions. In order to undertake such occurrence, the calendar ageing model should be paired with a cycle ageing model. The second paper of the series provides a detailed description of a counterpart ageing model, focussed on cycle battery operation [34]. The integration of both ageing models, within the context of defining a holistic view of lithium-ion degradation modelling is a challenging research task, and is the subject of ongoing research by the authors further extending the research presented here and in [34].

Finally, notice that the time and memory complexity of the GP is $O(n^3)$ and $O(n^2)$, respectively [23]. Therefore, the required computations rapidly become prohibitive within the context of increasing training datasets. Fortunately, a large number of approximation methods were proposed to overcome this problem [48,53], and the implementation of such solutions may be required once the training dataset becomes critically large. However, such issue should be mitigated regarding calendar ageing models, as i) the number of stress-factors is relatively low, leading to a reduced number of hyperparameters, and ii) the degradation in calendar operation is typically slower compared to a cycling use-case, and then larger time periods are needed to observe degradation, which reduces the number of training data extracted from real operation. This implies a limited growth of the training data and subsequently a restricted increase of computation time for the periodical update of the calendar ageing models. Nevertheless, this issue could be more critical for cycle ageing models, as commented in the Part B [34]. Lastly, it is noteworthy that the issue of the computational complexity of ageing models must be contrasted with the fact that Li-ion battery ageing is a relatively slow process, which does not require a rapid computation for the periodical update of the models.

## 9. Conclusions

In this paper, a calendar capacity loss model is developed based on the Gaussian Process framework. The model presents 0.31% $MAE_{\Delta Q}$ and 0.53% $MAE_{Q}$ average prediction errors for 30 cells operating between 25 °C-45 °C and 20–100% SOC storage conditions, using only 18 cells tested at 6 storage conditions for training.

This study illustrates the ability of GP-based ageing models to learn from the operating conditions progressively observed, increasing both accuracy and confidence of the model. The learning abilities of the GP models are validated at static and dynamic operating conditions. This makes the GP framework a suitable candidate to develop Li-ion ageing models able to evolve and improve their performances even after deployment in real application. Within this context, the suitability of isotropic kernel components to host the features corresponding to temperature and SOC storage conditions is also explored and validated. The sensitivity analysis shows that the developed model tends to assign a higher influence of the temperature variations on the capacity loss, compared to the SOC.

The model developed in this first paper is oriented to perform ageing predictions for applications implying large storage period of the Li-ion battery systems. Following an analogous method, the second paper of the series provides a detailed description of a counterpart ageing model, focussed on cycle battery operation.

## CRediT authorship contribution statement

**M. Lucu:** Data curation, Conceptualization, Methodology, Formal analysis, Software, Validation, Visualization, Writing - original draft. **E. Martinez-Laserna:** Investigation, Conceptualization, Methodology, Supervision, Writing - review & editing. **I. Gandiaga:** Conceptualization, Supervision, Writing - review & editing, Funding acquisition. **K. Liu:** Validation, Writing - review & editing. **H. Camblong:** Supervision, Writing - review & editing. **W.D. Widanage:** Writing - review & editing, Funding acquisition. **J. Marco:** Writing - review & editing, Funding acquisition.

## Declaration of Competing Interest

The authors declare that they have no known competing financial interests or personal relationships that could have appeared to influence the work reported in this paper.

## Acknowledgments

This investigation work was financially supported by ELKARTEK (CICe2018 - Desarrollo de actividades de investigación fundamental estratégica en almacenamiento de energía electroquímica y térmica para sistemas de almacenamiento híbridos, KK-2018/00098) and EMAITEK Strategic Programs of the Basque Government. In addition, the research was undertaken as a part of ELEVATE project (EP/M009394/1) funded by the Engineering and Physical Sciences Research Council (EPSRC) and partnership with the WMG High Value Manufacturing (HVM) Catapult.

Authors would like to thank the FP7 European project Batteries 2020 consortium (grant agreement No. 608936) for the valuable battery ageing data provided during the course of the project.

## Appendix A. Raw data obtained from experimental calendar ageing tests, and variability of the resulting capacity curves

Fig. A1, Table A1

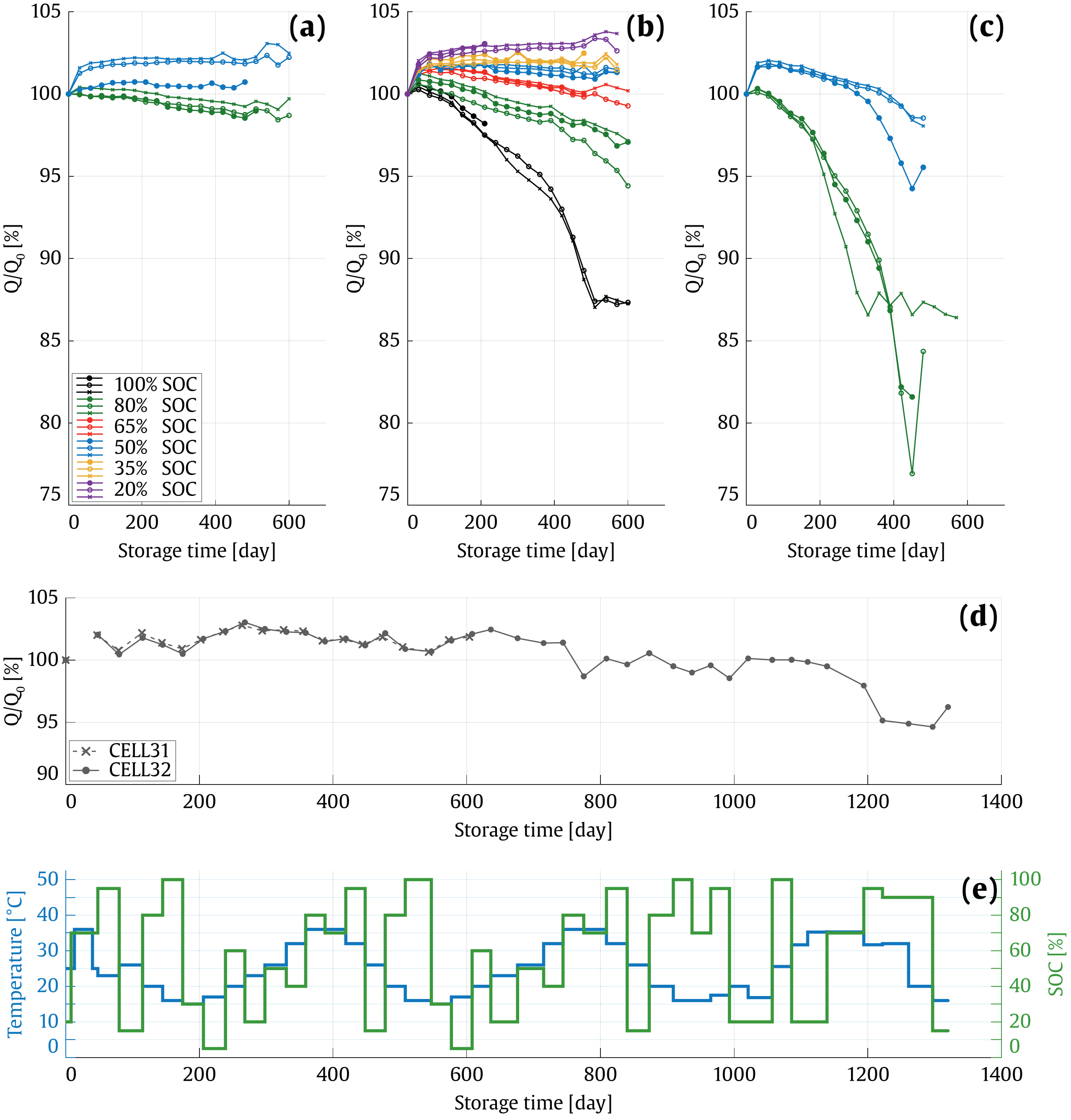

**Fig. A1.** Normalised (with initial value $Q_0$) capacity, obtained from the experimental static ageing tests at (a) 25 °C, (b) 35 °C and (c) 45 °C. (d) Normalised capacity, and (f) corresponding temperature and SOC storage conditions, obtained from the experimental dynamic ageing tests for CELL31. (e) Normalised capacity, and (g) corresponding temperature and SOC storage conditions, obtained from the experimental dynamic ageing tests for CELL32.

**Table A1**
Mean variance of the capacity curves, for the three cells tested at identical storage conditions (in [%$^2$]).

| Temperature [ °C] | SOC [%] | | | | | |
|---|---|---|---|---|---|---|
| | 100 | 80 | 65 | 50 | 35 | 20 |
| 25 | | 0.09 | | 0.67 | | |
| 35 | 0.05 | 0.42 | 0.02 | 0.03 | 0.04 | 0.03 |
| 45 | | 3.51 | | 0.95 | | |

## Appendix B. Results obtained with the models resulting from the different training cases

Table B1

**Table B1**
Results obtained with the models resulting from the different training cases, for the training, validation, and all the cells, in terms of ΔQ and Q.

| | Capacity loss (ΔQ) | | | | | | | | | Capacity (Q) | | | | | | | | |
|---|---|---|---|---|---|---|---|---|---|---|---|---|---|---|---|---|---|---|
| | Training | | | Validation | | | All | | | Training | | | Validation | | | All | | |
| | MAE | RMSE | CS | MAE | RMSE | CS | MAE | RMSE | CS | MAE | RMSE | CS | MAE | RMSE | CS | MAE | RMSE | CS |
| **CASE 1** | 0.37 | 0.45 | 85.27 | 0.66 | 0.75 | 90.74 | 0.60 | 0.69 | 89.61 | 0.72 | 0.78 | 61.54 | 2.02 | 2.30 | 59.75 | 1.75 | 1.98 | 60.12 |
| **CASE 2** | 0.31 | 0.39 | 89.93 | 0.53 | 0.61 | 84.08 | 0.46 | 0.54 | 85.90 | 0.639 | 0.703 | 69.66 | 1.33 | 1.54 | 58.87 | 1.11 | 1.28 | 62.22 |
| **CASE 3** | 0.26 | 0.33 | 94.51 | 0.37 | 0.44 | 99.49 | 0.30 | 0.37 | 96.42 | 0.47 | 0.53 | 83.03 | 0.64 | 0.76 | 89.04 | 0.53 | 0.62 | 85.31 |
| **CASE 4** | 0.27 | 0.34 | 94.87 | 0.28 | 0.33 | 100 | 0.27 | 0.33 | 96.28 | 0.50 | 0.55 | 81.99 | 0.37 | 0.41 | 100 | 0.46 | 0.51 | 86.95 |
| **CASE 5** | 0.28 | 0.34 | 95.31 | 0.26 | 0.30 | 100 | 0.27 | 0.33 | 96.28 | 0.49 | 0.54 | 83.55 | 0.44 | 0.53 | 94.44 | 0.48 | 0.54 | 85.80 |
| **CASE 6** | 0.25 | 0.32 | 95.75 | 0.38 | 0.43 | 100 | 0.27 | 0.33 | 96.19 | 0.44 | 0.49 | 84.99 | 0.41 | 0.42 | 100 | 0.43 | 0.48 | 86.54 |
| **CASE 7** | 0.26 | 0.33 | 96.19 | – | – | – | 0.26 | 0.33 | 96.19 | 0.45 | 0.50 | 86.54 | – | – | – | 0.45 | 0.50 | 86.54 |